\documentclass[letterpaper, 10pt, conference]{ieeeconf}

\IEEEoverridecommandlockouts                             

\usepackage{graphicx}
\usepackage{ascmac}

\usepackage{bm}
\usepackage{amsmath}
\usepackage{amssymb}
\usepackage{amsfonts}
\usepackage{color}
\usepackage{wrapfig}
\usepackage{url}
\usepackage{cite}
\usepackage{siunitx}
\usepackage{color}
\usepackage{comment}
\usepackage{caption}
\newcommand{\req}[1]{eq.~\eqref{#1}}
\newcommand{\rfig}[1]{Fig.~\ref{#1}}

\newtheorem{problem}{Problem}

\newtheorem{definition}{Definition}
\newtheorem{theorem}{Theorem}
\newtheorem{assumption}{Assumption}

\newtheorem{lemma}{Lemma}

\title{\LARGE \bf
Parameter Invariance Analysis of Moment Equations\\
Using Dulmage-Mendelsohn Decomposition
}

\author{Akito Igarashi and Yutaka Hori
\thanks{This work was supported in part by JSPS KAKENHI Grant Number 24K00911 and 24KK0267.}
\thanks{A. Igarashi and Y. Hori are with the Department
of Applied Physics and Physico-Informatics, Keio University, Kanagawa 223-8522 Japan. Correspondence should be addressed to Y. Hori. {\tt yhori@appi.keio.ac.jp}}
}

\begin{document}

\maketitle
\thispagestyle{empty}
\pagestyle{empty}

\begin{abstract}
Living organisms maintain stable functioning amid environmental fluctuations through homeostasis, a property that preserves a system's behavior despite changes in environmental conditions. To elucidate homeostasis in stochastic biochemical reactions, theoretical tools for assessing population-level invariance under parameter perturbations are crucial. In this paper, we propose a systematic method for identifying the stationary moments that remain invariant under parameter perturbations by leveraging the structural properties of the stationary moment equations. A key step in this development is addressing the underdetermined nature of moment equations, which has traditionally made it difficult to characterize how stationary moments depend on system parameters. To overcome this, we utilize the Dulmage-Mendelsohn (DM) decomposition of the coefficient matrix to extract welldetermined subequations and reveal their hierarchical structure. Leveraging this structure, we identify stationary moments whose partial derivatives with respect to parameters are structurally zero, facilitating the exploration of  fundamental constraints that govern homeostatic behavior in stochastic biochemical systems.
\end{abstract}

\section{Introduction}
\label{sec:introduction}
Homeostasis is the property by which organisms adapt to changes in external environments, enabling survival in highly uncertain conditions.
Experimental studies have begun to reveal the underlying structural mechanisms of biomolecular reactions that enable homeostatic behaviors \cite{muzzey2009, barkai1997, alon1999}. In this line of research, a recent study developed a synthetic biomolecular feedback system that is robust to disturbances \cite{AIF2019} based on the theoretical analysis of stochastic biomolecular reaction networks.
Mathematically, homeostasis can be described as the ability of the reaction network to maintain its state variables invariant to perturbations in parameters such as reaction rates, which may be affected by environmental factors such as temperature and osmolarity \cite{AIF2019, muzzey2009, khammash2022}.
To date, many efforts have been made to analyze the invariance based on deterministic models, specifically the rate equations. 
These include approaches that utilize reaction stoichiometry to identify subnetworks attaining perfect adaptation\cite{mochizuki2015, hirono2025}
and those employing sensitivity analysis based on the Jacobian matrix \cite{araujo2018,  blanchini2021}.
However, since intracellular reactions are inherently stochastic due to the low copy number of molecules, developing methods based on stochastic models is desirable for a more accurate analysis that better aligns with experimental observations. 

The governing equation for stochastic chemical reactions is the chemical master equation (CME), an infinite-dimensional linear differential equation for a discrete-state Markov process that poses challenges for sensitivity and invariance analysis \cite{mcquarrie1967, gillespie1992}. To address this, various numerical methods were proposed to evaluate the sensitivity of the density distributions of the copy numbers using the Fisher information matrix (FIM), employing Monte Carlo simulations or model approximations \cite{gunawan2005,komorowski2011}.
Meanwhile, moment-based approaches are amenable to analytic or algebraic characterization of stationary moments based on the moment equation \cite{sakurai2018, sakurai2022, khammash2022} and enabled rigorous analysis of invariant moments for a certain class of biomolecular networks. 
A notable advancement in this direction is the mathematical characterization of the antithetic integral feedback (AIF) network~\cite{AIF2016} as a system that achieves maximal robust perfect adaptation (MaxRPA), a property in which the stationary moments are robust to all but two parameter perturbations~\cite{gupta2022}. Yet, a general framework is still lacking for systematically identifying stationary moments that remain invariant under arbitrary parameter perturbations. Addressing this gap is essential for advancing the theoretical understanding of homeostasis beyond known motifs.

In this study, we propose a systematic method for analyzing the invariance of stationary moments with respect to arbitrary parameters by leveraging the structural information of the moment equations.
We utilize the Dulmage-Mendelsohn (DM) decomposition \cite{DM1958, DMsolver1990}, a graph-based matrix operation that extracts welldetermined subequations from a system of linear equations, to identify determinable variables in the highly underdetermined stationary moment equations and the dependency structure among the variables.
By linking this structure to unreachability analysis in directed graphs, we establish a 
computationally tractable algorithm to systematically identify stationary moments whose partial derivatives with respect to parameters are structurally zero.
The proposed method contributes to a broader understanding and application of homeostasis by explicitly accounting for the stochastic nature of biochemical reactions.

\noindent
\textbf{Notations:}
The symbol $\mathbb{P}(\cdot)$ is a probability mass function, and 
$\mathbb{E}[f(\cdot)]$ is the moment of the mass function defined by $\sum_{\bm{x}} f(\bm{x}) \mathbb{P}(\bm{x})$.
The symbol $\mathrm{supp}(\cdot)$ denotes the set of non-zero entries, \textit{i.e.,}
$\{(i,j) \mid (X)_{i,j} \neq 0\}$ for a matrix $X$ and $\{i \mid (\bm{x})_{i} \neq 0\}$ for a vector $\bm{x}$, respectively.
The expression 
$\bm{x}^{\bm{\alpha}}$ with $\bm{x}=[x_1, x_2, \ldots, x_n]^\top$ and $\bm{\alpha}=[\alpha_1, \alpha_2, \ldots, \alpha_n]^\top$ denotes a multivariate monomial of $x_1^{\alpha_1} x_2^{\alpha_2} \dots x_n^{\alpha_n}$.

\section{Problem Formulation}
In this section, we introduce a model describing the stochastic dynamics of chemical reactions and formulate the parameter invariance analysis problem.  

\subsection{Chemical Master Equations}
We consider chemical reaction networks $\Gamma$ consisting of $n$ molecular species $X_i \:(i = 1, 2, \dots, n)$ and $m$ reaction species $R_j \:(j = 1, 2, \dots, m)$. 
In stochastic chemical reactions, the state of the reaction networks $\Gamma$ is characterized by the copy number of molecular species $X_i$, and its evolution is modeled by the Markov process governed by stochastic 
changes in 
the copy numbers by reactions $R_j$. 

To formally define this stochastic process, let $x_i$ denote the copy number of the molecular species $X_i$, and define the state of the reaction network $\Gamma$ at time $t$ by $\bm{x}(t) = [x_1, x_2, \ldots, x_n]^\top \in \mathbb{N}_0^n$. 
Reaction species $R_j$ is characterized by a stoichiometry vector $\bm{s}_j \in \mathbb{Z}^n$, which describes the change in $\bm{x}$ upon the occurrence of the reaction, and a propensity function $w_j (\bm{x})$ representing the probability of its occurrence per unit time.
The reaction $R_j$ is one of the elementary reactions, and the propensity function $w_j(\bm{x})$ is linear with respect to $\theta_j$ and a polynomial in $\bm{x}$ with a maximum degree of two~\cite{khammash2022}.

The stochastic chemical reactions are then modeled as the evolution of the copy number distribution of the state $\bm{x}$ at time $t$ denoted by $\mathbb{P}\left(\bm{x}| \bm{\theta},t,\bm{x}_0\right)$, for a given initial state $\bm{x}_0$, \textit{i.e.,} $\bm{x}_0 := \bm{x}(0)$. 
The governing equation of the dynamics is called 
the Chemical Master Equation (CME) \cite{mcquarrie1967, gillespie1992}. 
Omitting $t$ and $\bm{x}_0$ in $\mathbb{P}\left(\bm{x}| \bm{\theta}, t,\bm{x}_0\right)$ to avoid notational clutter, the CME can be expressed as 
\begin{align}
    \label{CME}
\!\!\!\!\!
    \frac{d \mathbb{P}(\bm{x} | \bm{\theta})}{dt} \!=\! & \sum_{j=1}^{m} \{w_{j}(\bm{x}\!-\!\bm{s}_{j}) \mathbb{P}(\bm{x}\!-\!\bm{s}_{j} | \bm{\theta})
    \!-\! w_{j}(\bm{x})\mathbb{P}(\bm{x} | \bm{\theta})\}\!\!\!\!
\end{align}
for each $\bm{x} \in \mathbb{N}_0^n$, where $\bm{\theta} := [\theta_1, \theta_2, \ldots, \theta_m]^\top \in \mathbb{R}_+^m$. 
It should be noticed that the CME is a linear but an infinite dimensional ordinary differential equation (ODE), for which analytic solutions are rarely available.

\subsection{Problem statement}
In the design and analysis of stochastic reaction networks, one is often interested in understanding how the copy number distributions $\mathbb{P}(\bm{x}|\bm{\theta})$ depend on the rate parameters $\bm{\theta}$, which can be influenced by environmental changes such as temperature and osmolarity\cite{khammash2022}. 
In this work, we examine these dependencies through the moments of the copy number distribution $\mathbb{E}[\bm{x}^\alpha|\bm{\theta}]$, which capture essential statistics such as the mean and the variance. 
More specifically, our focus is on the stationary moments, and we investigate whether certain stationary moments remain invariant under changes in the rate parameters $\bm{\theta}$, as such invariance is closely related to the homeostatic behavior of biomolecular systems.

To formally define the problem, we introduce the following assumption, following literature \cite{sakurai2022}.

\begin{assumption}
\label{assm: moment}
For any rate parameters $\bm{\theta}$ and initial state $\bm{x}_0$, (i) the stationary solution of the CME (\ref{CME}) exists, 
and (ii) its associated Markov process is non-explosive. 
Moreover, (iii) all moments of the stationary distributions are finite.
\end{assumption}

This assumption ensures the existence of a stationary distribution and stationary moments of molecular copy numbers, which is closely related to the ergodicity of the underlying Markov process \cite{briat2023}. 
The ergodicity of stochastic chemical reaction networks can be verified using tractable optimization algorithms \cite{gupta2014}.
In what follows, we denote the stationary moments by $\mu_{\bm{\alpha}}(\bm{\theta}) := \mathbb{E}[\bm{x}^{\bm{\alpha}}|\bm{\theta}]$, where $\bm{\alpha} \in \mathbb{N}_0^m$ is a multi-index specifying the moment (see Notations in Section \ref{sec:introduction}). 
Then, the problem is formally stated as follows.

\noindent
\begin{problem}
\textit{
For a given chemical reaction network $\Gamma$ and one of the reaction species $R_k$,  
assume that Assumption \ref{assm: moment} holds. 
Define the variation of the parameter associated with the reaction species $R_k$ around $\bm{\theta}$ by
\begin{align}
\Delta_k(\bm{\theta}) := \{ \bm{\theta} + \phi \bm{\delta}_k \mid  \phi \in \mathbb{R}, \bm{\theta} + \phi \bm{\delta}_k \in \mathbb{R}_+^m\},
\end{align}
where $\bm{\delta}_k$ is the $k$-th standard basis of $\mathbb{R}^m$ with the $k$-th entry being 1 and the others being 0.
Under this parameter variation, find a set of invariant stationary moments 
\begin{align}
\label{problem_M}
\mathcal{M}_k := \left\{ \bm{\alpha}  \ \middle| \ 
\begin{aligned}
&{\mu}_{\bm{\alpha}}(\bm{\theta}+\delta\bm{\theta}) = {\mu}_{\bm{\alpha}}(\bm{\theta})
\\
& \forall \bm{\theta} \in \mathbb{R}^m_+,\ \forall \delta\bm{\theta} \in \Delta_k(\bm{\theta})
\end{aligned}
\right\}
.
\end{align}
}
\end{problem}

\section{Analysis of Invariant Moments}
In this section, we present a systematic method for identifying the invariant stationary moments $\mathcal{M}_k$. 
Our approach analyzes the sensitivity of stationary moments with respect to the parameters and derives an algebraic equation that characterizes this sensitivity. We then develop an algorithm that systematically explores the structural properties of this equation to identify the invariant stationary moments.

\subsection{Overview of the proposed approach}
\label{sec:overview}
Let $\partial_k \mu_{\bm{\alpha}} (\bm{\theta})$ denote the partial derivative of the stationary moment with respect to $\theta_k$ defined as
\begin{align}
\label{partial}
\partial_k \mu_{\bm{\alpha}}(\bm{\theta}) := \frac{\partial \mu_{\bm{\alpha}}}{\partial \theta_k} \bigg|_{\bm{\theta}}.
\end{align} 
By definition, the stationary moment $\mu_{\bm{\alpha}}(\bm{\theta})$ is invariant under the change in the parameter $\theta_k$, \textit{i.e.,} $\bm{\alpha} \in \mathcal{M}_k$ holds, if and only if 
$\partial_k \mu_{\bm{\alpha}} (\bm{\theta}) = 0$ for all $\bm{\theta}$. 
In other words, our problem reduces to identifying the moments that remain invariant due to the structural properties of the system, independent of the specific parameter values $\bm{\theta}$.

To explore such moments, we consider the stationary moment equation. 
Specifically, we first obtain the linear equation for the stationary distribution by setting the left-hand side of \req{CME} to zero. Next, we multiply these linear equation by $\bm{x}^{\bm{\alpha}}$ and sum over all $\bm{x} \in \mathbb{N}^n_0$ to derive the moment equation
\cite{sakurai2018}
\begin{align}
    \label{ME}
    \sum_{\bm{\beta}}\sum_{j=1}^{m} 
    \theta_j (a^j_{\bm{\alpha}, \bm{\beta}} 
    \mu_{\bm{\beta}} (\bm{\theta}) 
    + b^j_{\bm{\alpha}}) = 0,
\end{align}
where $a^j_{\bm{\alpha}, \bm{\beta}}$ and $b^j_{\bm{\alpha}}$ are coefficients determined by the stoichiometry vectors $\bm{s}_j$ and propensity functions $w_j(\bm{x})$. 
Equation \eqref{ME} is a set of linear equations that the stationary moment $\mu_{\bm{\alpha}} (\bm{\theta})$ must satisfy,  
and the coefficient $a^j_{\bm{\alpha}, \bm{\beta}}$ represents the interaction between the stationary moments $\mu_{\bm{\alpha}}$ and $\mu_{\bm{\beta}}$ through reaction $R_j$, while $b^j_{\bm{\alpha}}$ is a constant term arising from zero-th order reactions or conserved quantities of molecular copy number. 
Taking the derivative of eq. \eqref{ME} with respect to ${\theta}_k$, we have 
\begin{align}
\label{deriv-ME}
    &\sum_{\bm{\beta}}
    \left\{
    a^k_{\bm{\alpha}, \bm{\beta}} 
    \mu_{\bm{\beta}} (\bm{\theta}) 
    + b^k_{\bm{\alpha}}
    +
    \sum_{j=1}^{m} 
    \theta_j a^j_{\bm{\alpha}, \bm{\beta}} 
    \partial_k \mu_{\bm{\beta}} (\bm{\theta})
    \right\}=0, 
\end{align}
which is a linear equation with respect to the moments and their partial derivatives with respect to $\theta_k$. 
In typical cases where the reaction network $\Gamma$ includes bimolecular reactions, the stationary moment equation is underdetermined since each moment equation depends on higher order moments. This dependence leads to an analytically intractable infinite dimensional equation, and thus, it is not straightforward to analyze the parameter dependence of the moments. 

To address this challenge, we present an approach that utilizes the structural information of these algebraic equations to identify the invariant moments satisfying $\partial_k \mu_{\bm{\alpha}} (\bm{\theta}) = 0$ for all $\bm{\theta} \in \mathbb{R}_+^m$.
To this end, we first replace the variables (moments) in eq. \eqref{ME} and eq. \eqref{deriv-ME} with independent variables and define the following linear equations
\begin{align}
    \Phi(\bm{\theta}) \bm{\nu}_\Phi &= \bm{b}_\Phi(\bm{\theta}),        \label{perturbed_equation}
\end{align}
where 
\begin{align}
    \!\!\!\Phi(\bm{\theta}) \!:=\!
\begin{bmatrix}
    A(\bm{\theta}) \!\!\!&  \!\!\!A_k\\
    O  \!\!\!& \!\!\! A(\bm{\theta})
\end{bmatrix},
\bm{\nu}_\Phi \!:=\!
\begin{bmatrix}
         \partial_k \bm{\nu}\\
         \bm{\nu}
\end{bmatrix},
\bm{b}_\Phi(\bm{\theta})\! :=\!
-
\begin{bmatrix}
         \bm{b}_k\\
         \bm{b}(\bm{\theta})
\end{bmatrix}\!
\end{align}
with $ A(\bm{\theta}) := \sum_{j = 1}^m \theta_j A_j $, $ \bm{b}(\bm{\theta}) := \sum_{j = 1}^m \theta_j \bm{b}_j $, 
$ A_j \in \mathbb{R}^{M \times N} $ and $ \bm{b}_j \in \mathbb{R}^M\:(j = 1, 2, \dots, m)$ are matrices and vectors containing the coefficients $ a_{\bm{\alpha}, \bm{\beta}}^j $ and $ b_{\bm{\alpha}}^j $, respectively, and   
the vectors $ \bm{\nu} $ and $\partial_k \bm{\nu}$ denote independent variables associated with the stationary moments and their derivatives.
For simplicity, we make the following assumptions that eliminate redundant equations and unnecessary variables.

\begin{assumption}
    \label{assumption2}
For all $\bm{\theta} \in \mathbb{R}^{m}_+$, the matrix $A(\bm{\theta})$ in eq. \eqref{perturbed_equation} is full row rank, and at least one entry of 
each column of $A(\bm{\theta})$ is non-zero.
\end{assumption}

The full row rank condition implies that the moment equation does not have redundant equations, which can be systematically eliminated by utilizing the basis vectors common to the left null spaces of the augmented coefficient matrices $[A_j \ \bm{b}_j]\: (j = 1,2,\dots,m)$. 
The latter half of the assumption ensures that the vectors $\bm{\nu}_\Phi$ do not contain unused variables.

In the rest of this section, we illustrate the overview of the proposed approach using a simple example.

\noindent
\textbf{Example:} 
Consider the reaction network with molecular sequestration $\Gamma$, where the stoichiometric vectors $\bm{s}_j$ and propensity functions $w_j(\bm{x})$ are defined in Table \ref{AIF_model_def}.
\begin{table}[tb]
  \small
  \caption{Definitions of the chemical reaction network in Example}
  \label{AIF_model_def}
  \centering
  \begin{tabular}{ccc}
    \hline
    $R_j$ & Stoichiometry vector $\bm{s}_j$ & Propensity function $w_j(\bm{x})$ \\
    \hline 
     $R_1$ & $[1, 0, 0]^\top$ & $\theta_1$ \\
     $R_2$ & $[0, 1, 0]^\top$ & $\theta_2 x_1$ \\
     $R_3$ & $[0, 0, 1]^\top$ & $\theta_3 x_2$ \\
     $R_4$ & $[-1, 0, -1]^\top$ & $\theta_4 x_1 x_3$ \\
     $R_5$ & $[0, -2, 0]^\top$ & ${\theta_5 x_2 (x_2 - 1)}/2$ \\
    \hline
  \end{tabular}
\end{table}
\begin{figure}[t] 
  \begin{center}
    \includegraphics[width=75mm]{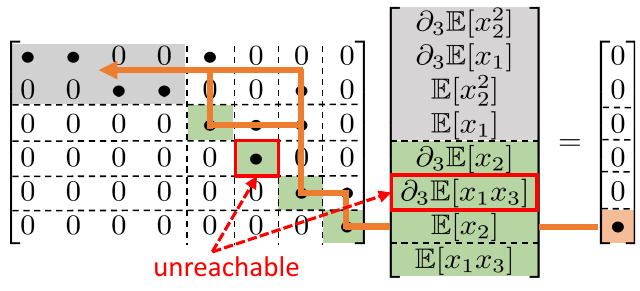}
  \end{center}
  \caption{Identification of invariant moments based on the row and column permutation 
  of eq. \eqref{example_moment}.
  The rearranged equation shows $\partial_3 \mathbb{E}[x_1 x_3] = 0$ 
  for all $\bm{\theta}$.}
  \label{concept}
\end{figure}
Our goal is to find invariant moments under perturbations in the parameter $\theta_3$ of the reaction $R_3$.
When the moment equations are truncated up to $\sum^n_{i = 1} \alpha_i = 1$, eq.~\eqref{perturbed_equation}
is given by 
\begin{align}
    \label{example_moment}
    \!\!\!\!\begin{bmatrix}
\!0 &\!\!\! 0 &\!\!\!\! 0 &\!\!\!\!\! -\theta_4 &\!\!\! 0 &\!\!\! 0 &\!\!\!\! 0 &\!\!\!\!\!\! 0\\
\theta_2 &\!\!\! \theta_5 &\!\!\!\! -\theta_5 &\!\!\!\!\! 0 &\!\!\! 0 &\!\!\! 0 &\!\!\!\! 0 &\!\!\!\!\!\! 0 \\
\!0 &\!\!\! \theta_3 &\!\!\!\! 0 &\!\!\!\!\! -\theta_4 &\!\!\! 0 &\!\!\! 1 &\!\!\!\! 0 &\!\!\!\!\!\! 0 \\
\!0 &\!\!\! 0 &\!\!\!\! 0 &\!\!\!\!\! 0 &\!\!\! 0 &\!\!\! 0 &\!\!\!\! 0 &\!\!\!\!\!\! -\theta_4 \\
\!0 &\!\!\! 0 &\!\!\!\! 0 &\!\!\!\!\! 0 &\!\!\! \theta_2 &\!\!\! \theta_5 &\!\!\!\! -\theta_5 &\!\!\!\!\!\! 0 \\
\!0 &\!\!\! 0 &\!\!\!\! 0 &\!\!\!\!\! 0 &\!\!\!\! 0 &\!\!\! \theta_3 &\!\!\!\! 0 &\!\!\!\!\!\! -\theta_4 \\
\end{bmatrix}\!\!\!
\begin{bmatrix}
{\partial_3 \mathbb{E}}[x_1]\!\!\\
{\partial_3 \mathbb{E}}[x_2]\!\!\\
{\partial_3 \mathbb{E}}[x^2_2]\!\!\\
{\partial_3 \mathbb{E}}[x_1x_3]\!\\
{\mathbb{E}}[x_1]\!\!\\
{\mathbb{E}}[x_2]\!\!\\
{\mathbb{E}}[x^2_2]\!\!\\
{\mathbb{E}}[x_1x_3]\!\!
\end{bmatrix}
\!\!=\!\!
\begin{bmatrix}
    0\\
    0\\
    0\\
    \!-\theta_1\!\\
    0\\
    0
\end{bmatrix}\!, \!
\end{align}
where $\partial_3 \mathbb{E}[\cdot]$ represents the derivative of the moment with respect to $\theta_3$. 
This equation is underdetermined, consisting of eight variables and six equations, making it difficult to directly identify the invariant stationary moments by solving the equation. 

Hence, we reformulate eq. \eqref{example_moment} to extract the dependency between the variables. Specifically, 
by permuting the rows and columns of the coefficient matrix into an upper block-triangular form and rearranging the corresponding vectors, eq. \eqref{example_moment} is transformed into the structure shown in Fig.~\ref{concept}, where non-zero elements are indicated by $\bullet$. 
In Fig. \ref{concept}, the equation is separated into square diagonal blocks highlighted in green and the remaining rectangular block 
shown in gray. Therefore, as indicated by the orange arrow, backward substitution reveals that the third diagonal block  
from the bottom is independent of the right-hand side, implying that $\mathbb{E}[x_1 x_3]$ is the invariant moment satisfying $\partial_3\mathbb{E}[x_1 x_3] = 0$ for all $\bm{\theta} \in \mathbb{R}^m_+$. 
$\hfill \Box$

A key idea in this example is the rearrangement of eq. \eqref{example_moment} into the welldetermined and the underdetermined subequations shown in Fig. \ref{concept} and the subsequent backward substitution. 
In the next subsection, we show that such a transformation can be carried out systematically by using 
the Dulmage-Mendelsohn (DM) decomposition \cite{DM1958} and that the invariant moments can be found by the reachability analysis of a graph associated with the rearranged equation.

\subsection{Dulmage-Mendelsohn decomposition}
The DM decomposition is a graph theoretic method that provides the finest possible block-triangular decomposition of a matrix, separating its structurally welldetermined part from a given matrix \cite{DM1958, DMsolver1990}. 
Specifically, for each parameter value $\bm{\theta}$, we define a bipartite graph $\mathcal{B}(A(\bm{\theta}))$ associated with the matrix $A(\bm{\theta})$, where one set of nodes corresponds to the rows (row nodes) and the other to the columns (column nodes), and an edge connects a row node and a column node if the corresponding entry of $A(\bm{\theta})$ is non-zero. 
The following lemma follows directly from Theorems 2.1 and 2.2 of \cite{DMsolver1990} with Assumption \ref{assumption2}.

\begin{lemma} 
\label{lemma1}
Consider a matrix $A(\bm{\theta})$ satisfying Assumption \ref{assumption2}, and the bipartite graph $\mathcal{B}(A(\bm{\theta}))$. Let $M$ be a maximum matching of $\mathcal{B}(A(\bm{\theta}))$. 
Suppose that some nodes in $\mathcal{B}(A(\bm{\theta}))$ are not reachable from unmatched column nodes via alternating paths with respect to $M$.
Then, there exists permutation matrices $P$ and $Q$ such that 
\begin{equation}
    \label{DM}
    P {A}(\bm{\theta}) Q = 
    \begin{bmatrix}
        {A}_u (\bm{\theta}) & {A}_{ud}(\bm{\theta}) \\
        O & {A}_d(\bm{\theta})
    \end{bmatrix},
\end{equation}
where ${A}_d \in \mathbb{R}^{N_d \times N_d}$ is a block upper-triangular matrix with $\eta$ diagonal blocks satisfying $\mathrm{rank}({A}_d) = N_d$, and ${A}_u \in \mathbb{R}^{(M - N_d) \times (N - N_d)}$ satisfies $\mathrm{rank}({A}_u) = M - N_d$.
Among such transformations, the partition maximizing $\eta$ is unique, and each diagonal block in $A_d$ is uniquely determined up to row and column permutations within that block.
\end{lemma}

The permutation matrices $P$ and $Q$ are obtained by computing the maximum matching of the bipartite graph, followed by a strongly connected component decomposition of the square block $A_d$. 
Polynomial-time algorithms are known for performing these operations \cite{DMsolver1990}.
Lemma 1 enables the extraction of the moments and their derivatives that can be fully determined from the underdetermined moment equations when the bipartite graph $\mathcal{B}(A(\bm{\theta}))$ satisfies the assumption in Lemma 1. 
The next lemma shows that the maximal welldetermined block of $\Phi(\bm{\theta})$ in eq. (\ref{perturbed_equation}) can be obtained using $A_d(\bm{\theta})$.

\begin{lemma}
\label{lemma2} 
Suppose 
the matrix $A(\bm{\theta})$ satisfies Assumption \ref{assumption2} and admits 
the decomposition given in  
eq. \eqref{DM}. There exist row and column permutations that transform $\Phi(\bm{\theta})$ into the upper-triangular block matrix  
    \begin{align}
    \label{eq:Psi}
        \Psi(\bm{\theta})
        =
        \begin{bmatrix}
            {A}_{u1}(\bm{\theta})&{A}_{u2}(\bm{\theta})&{A}_{u3}(\bm{\theta})\\
            O&{A}_d(\bm{\theta})&{A}^k_d\\
            O&O&{A}_d(\bm{\theta})
        \end{bmatrix},
\end{align}
where ${A}_d \in \mathbb{R}^{N_d \times N_d}$ is a block-triangular matrix with $\eta$ diagonal blocks defined in eq. \eqref{DM}, and ${A}_d^k \in \mathbb{R}^{N_d \times N_d}, {A}_{u1}\in \mathbb{R}^{2(M-N_d) \times 2(N-N_d)}, {A}_{u2} \in \mathbb{R}^{2(M-N_d) \times N_d}$ and ${A}_{u3} \in \mathbb{R}^{2(M-N_d) \times N_d}$ are real-valued matrices defined accordingly. 
Moreover, this decomposition maximizes the diagonal blocks among all row and column permutations of $\Phi$, and each diagonal block of $\Psi$ is unique up to row and column permutations within the block.
\end{lemma}

The proof of this lemma is shown in Appendix \ref{app.1}. 
Lemma \ref{lemma2} implies that eq. \eqref{perturbed_equation} can be systematically rearranged by the DM decomposition to extract the square submatrix corresponding to the welldetermined equations. 
In the next subsection, we build upon this result by developing a graph-based algorithm that generalizes the backward substitution illustrated in the example in Section \ref{sec:overview}. 
In particular, the proposed algorithm relies solely on the structural information of $\Psi(\bm{\theta})$ and does not require the DM decomposition for each parameter instance, enabling efficient  identification of invariant moments for all $\bm{\theta} \in \mathbb{R}^m_+$.

\subsection{Graph-based identification of invariant moments}
\label{sec:graph}

We first define a \textit{saturated} matrix and vector as follows. 
\begin{definition}
\label{def1}
A matrix ${X}(\bm{\theta})$ and a vector $\bm{x}(\bm{\theta})$ parameterized by $\bm{\theta} \in \mathbb{R}_+^m$ are said to be saturated if its sparsity pattern equals the union of all possible supports,
that is, 
$\mathrm{supp}(X(\bm{\theta})) = \bigcup_{\bm{\theta}_o \in \mathbb{R}_+^m} \mathrm{supp}(X(\bm{\theta}_o))$ and $
\mathrm{supp}(\bm{x}(\bm{\theta})) = \bigcup_{\bm{\theta}_o \in \mathbb{R}_+^m} \mathrm{supp}(\bm{x}(\bm{\theta}_o))$, where $\mathrm{supp}(\cdot)$ is defined in Section \ref{sec:introduction}.
\end{definition}

\noindent
It is important to note that the support (sparsity pattern) is identical across all saturated matrices, and likewise for all saturated vectors.
In practice, $A(\bm{\theta})$ and $\bm{b}(\bm{\theta})$ in eq. \eqref{perturbed_equation} are saturated for most choices of $\bm{\theta}$ as their non-zero patterns are preserved unless the entries vanish due to the cancellation of the parameters. 
In such typical cases, the sparsity patterns of the saturated matrix $A(\bm{\theta})$ and vector $\bm{b}(\bm{\theta})$ are characterized by  constant matrices and vectors as
$\bigcup_{j} \mathrm{supp}(A_j)$ and 
$\bigcup_{j} \mathrm{supp}(\bm{b}_j)$, respectively.

Based on this definition, we show a pivotal lemma that leads to the identification of invariant moments based solely on structural information of the equation.
\begin{lemma}
\label{lemma3}
Suppose the matrix $A(\bm{\theta})$ satisfies Assumption \ref{assumption2}.  
Then, for any $\bm{\theta} \in \mathbb{R}_+^m$ such that $A(\bm{\theta})$ is saturated and admits the decomposition \eqref{DM}, 
there exist common permutation matrices $P$ and $Q$ in eq. \eqref{DM} that transform $\Phi(\bm{\theta})$ into $\Psi(\bm{\theta})$. Moreover, the matrix $\Psi(\bm{\theta})$ obtained in this way is saturated. 
\end{lemma}
\begin{proof}
Since the bipartite graph $\mathcal{B}(A(\bm{\theta}))$ is determined by the support of $A(\bm{\theta})$, the same $P$ and $Q$ apply to  
eq. (\ref{DM}) for all $\bm{\theta}$ that makes  $A(\bm{\theta})$ saturated. By Lemma \ref{lemma2}, these $P$ and $Q$ transform $\Phi(\bm{\theta})$ into $\Psi(\bm{\theta})$. 
The matrix $\Psi(\bm{\theta})$ is  saturated since permutations only relabel the rows and columns, and $\Phi(\bm{\theta})$ is saturated when $A(\bm{\theta})$ is saturated.
\end{proof}

This lemma implies that a single computation of the DM decomposition for some saturated matrix $A(\bm{\theta})$ suffices to obtain the permutation matrices $P$ and $Q$ leading to the coefficient matrix $\Psi(\bm{\theta})$ of the rearranged equation. 
More explicitly, we define a linear equation  that is obtained by rearranging eq. \eqref{perturbed_equation} using the common permutation matrices in Lemma \ref{lemma3} as 
\begin{align}
    \label{eq:DM_moment}
    \Psi (\bm{\theta})\bm{\nu}_{\Psi} = \bm{b}_{\Psi}(\bm{\theta}), 
\end{align}
 where $\bm{\nu}_\Psi = [{\bm{\nu}}_{0}, {\bm{\nu}}_{1}, \ldots, {\bm{\nu}}_{2\eta}]^\top \in \mathbb{R}^{2N}$ and $\bm{b}_\Psi(\bm{\theta}) = [{\bm{b}}_{0}, {\bm{b}}_{1}, \ldots, {\bm{b}}_{2\eta}]^\top \in \mathbb{R}^{2M}$ are the vectors obtained by reordering the entries of $\bm{\nu}_\Phi$ and $\bm{b}_\Phi(\bm{\theta})$, respectively, according to the transformation from the matrix $\Phi(\bm{\theta})$ to the matrix $\Psi(\bm{\theta})$. 
The sub-vectors ${\bm{\nu}}_{i}$ and $\bm{b}_{i}$ $(i=0,1,\ldots,2\eta)$ are defined in correspondence with the $i$-th diagonal block of $\Psi$, where the subscript 0 refers to the upper-left block ${A}_{u1}$ of $\Psi$. 
By construction,  
${\bm{\nu}}_{i+\eta}$ and ${\bm{\nu}}_{i} \: (i = 1,2,\dots,\eta)$ represent the stationary moments and their partial derivatives, respectively.

In what follows, we show that the identification of invariant moments reduces to a reachability analysis on a graph induced by the structure of eq. \eqref{eq:DM_moment}. To this end, we define 
a graph $\mathcal{G}(\mathcal{V}, \mathcal{E})$ with node set $\mathcal{V} := \{V_1, V_2, \ldots, V_{2\eta}\}$ and edge set $\mathcal{E}$ that encodes the structure of the sparsity pattern of $\Psi$, where 
each node $V_i \in \mathcal{V}$ corresponds to each variable ${\bm{\nu}}_i$, and an edge $(V_q, V_p)$ is introduced from $V_q$ to $V_p$ if the corresponding block in $\Psi$ is non-zero. 
Specifically, the edge set is defined by 
\begin{align}
\mathcal{E} := 
\{
(V_q, V_p) \in \mathcal{V} \times \mathcal{V} \mid 
\Psi[\mathcal{R}_p, \mathcal{C}_q] \neq O
\}, 
\end{align}
where $\mathcal{R}_i$ and $\mathcal{C}_i$ represent the row and column indices corresponding to the $i$-th diagonal block of the matrix $\Psi$ $(i=0,1,\ldots,2\eta)$, and $\Psi[\mathcal{R}_p, \mathcal{C}_q]$ is the $(p,q)$ block of $\Psi$ specified by the row and column indices $\mathcal{R}_p$ and $\mathcal{C}_q$.
In general, the topology of the graph $\mathcal{G}(\mathcal{V}, \mathcal{E})$ is uniquely determined for all $\bm{\theta}$ such that $A(\bm{\theta})$ is saturated since $\Psi(\bm{\theta})$ is also saturated due to Lemma \ref{lemma3}, and the sparsity pattern of saturated matrices are identical.

The following theorem shows a graph-based analysis method for 
identifying the invariant moments.

\begin{theorem}
\label{theorem_1}
Suppose Assumption~\ref{assumption2} holds. Let $\bm{\theta}^*$ be the parameter such that $A(\bm{\theta}^*)$ admits the decomposition \eqref{DM}, and $A(\bm{\theta}^*)$ and $\bm{b}(\bm{\theta}^*)$ are saturated. Define the graph $\mathcal{G}(\mathcal{V}, \mathcal{E})$ induced by eq.~\eqref{eq:DM_moment} with $\bm{\theta}^*$, and let the set of source nodes by $\mathcal{S} := \{V_i \in \mathcal{V} \mid {\bm{b}}_i \neq \bm{0}\} (\subseteq \mathcal{V})$. 
Then, ${\bm{\nu}}_i = \bm{0}$ in eq. \eqref{eq:DM_moment} holds for any $\bm{\theta} \in \mathbb{R}^m_+$ if $V_i$ is unreachable from all source nodes $\mathcal{S}$.
\end{theorem}

\begin{proof}
The proof is divided in two parts. 
(i) For a fixed $\bm{\theta}^*$, consider eq. \eqref{eq:DM_moment}. The equation corresponding to the row set $\mathcal{R}_i$ is 
\begin{align}
    \label{theorem1_eq1}
            \Psi[\mathcal{R}_{i}, \mathcal{C}_i] {\bm{\nu}}_{i} + \sum_{q \in \mathcal{N}_i} \Psi[\mathcal{R}_{i}, \mathcal{C}_q] {\bm{\nu}}_{q} = {\bm{b}}_i
\end{align}
where $\mathcal{N}_i := \{j \mid (V_j, V_i) \in \mathcal{E}\}$. 
It follows that ${\bm{\nu}}_i = \bm{0}$ if ${\bm{\nu}}_q$ for $q \in \mathcal{N}_i$ and $\bm{{b}}_i$ are zero $(\star)$.
Since $V_i$ is unreachable from the source nodes, 
${\bm{b}}_q = 0$ for every equation associated with upstream nodes $V_q$ along the directed paths leading to $V_i$.
Moreover, the equations for the most upstream node $V_j$ reduces to 
$\Psi[\mathcal{R}_{j}, \mathcal{C}_{j}] {\bm{\nu}}_j = \bm{0}$, which implies ${\bm{\nu}}_j = \bm{0}$ as $\Psi[\mathcal{R}_{j}, \mathcal{C}_{j}]$ is a full rank square matrix
due to Assumption 2. 
Thus, by recursively applying $(\star)$ from the most upstream node to $V_i$, we have ${\bm{\nu}}_i = 0$.

(ii) Next, we show that the same result holds for any $\bm{\theta} \in \mathbb{R}_+^m$. Consider $\bm{\theta}$ such that $A(\bm{\theta})$ and $\bm{b}(\bm{\theta})$ are saturated. Then, the graph and the source nodes induced by $\Psi(\bm{\theta})$ and $\bm{b}(\bm{\theta})$ are identical to those induced by $\Psi(\bm{\theta}^*)$ and $\bm{b}(\bm{\theta}^*)$ since $\Psi(\bm{\theta})$ is saturated by Lemma \ref{lemma3}, and 
saturated matrices and vectors have
the same sparsity pattern regardless of the parameter $\bm{\theta}$. 
Thus, by the same argument as in (i), ${\bm{\nu}}_i = \bm{0}$ holds. 
For other $\bm{\theta}$, where either $A(\bm{\theta})$ or $\bm{b}(\theta)$ is not saturated, some entries of $\Psi(\bm{\theta})$ or $\bm{b}(\theta)$ that were not zero in the saturated case become zero. 
As a result, the associated graph and the source nodes are obtained by removing the corresponding edges and source nodes from the graph $\mathcal{G}(\mathcal{V}, \mathcal{E})$ for the saturated case. 
As these are subsets of those in the saturated case, unreachability is preserved. 
Thus, the same argument as in (i) leads to ${\bm{\nu}}_i = \bm{0}$. 
\end{proof}

Theorem \ref{theorem_1} shows that, based on the idea of backward substitution, the analysis of parameter invariance of stationary moments can be reduced to a reachability problem in a directed graph.
The Floyd-Warshall algorithm provides an efficient way to compute the reachability from multiple source nodes $\mathcal{S}$ in $O(|\mathcal{V}|^3)$ time with respect to the number of nodes $|\mathcal{V}|$\cite{floyd1962}.

\par
\noindent
\textbf{Remark:} 
The proposed method does not guarantee identification of all invariant moments $\mathcal{M}_k$. This limitation stems from two main factors: 
(i) some invariant moments may not be identified if they depend on higher order terms omitted by the truncation of moment equation, (ii) invariance arising from dependencies among matrix entries cannot be detected as the proposed approach uses only the sparsity pattern.

\section{Application example}
\begin{figure}[t] 
  \begin{center}
    \includegraphics[width=85mm]{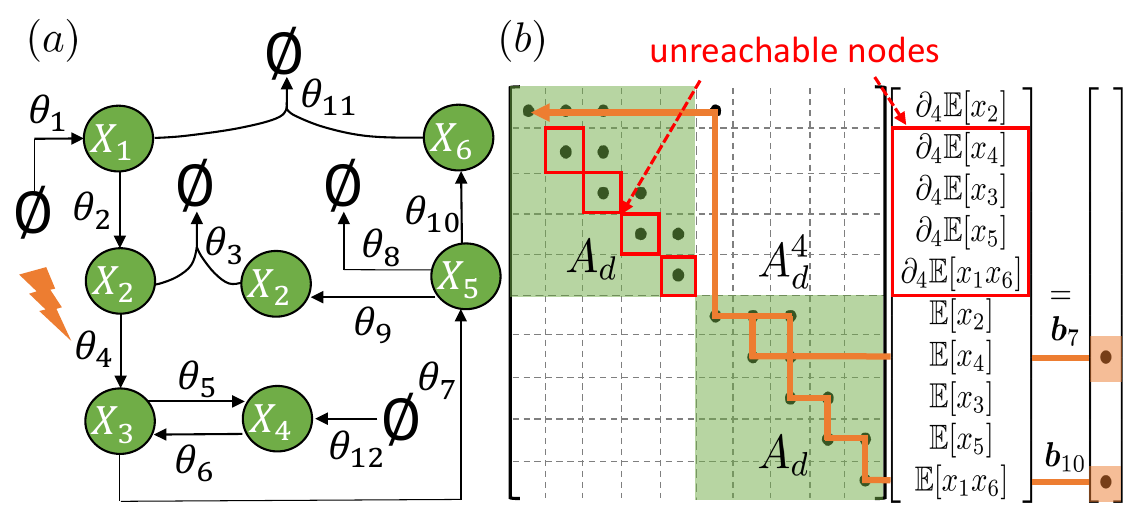}
  \end{center}
  \caption{Application example. (a) a network of feedback reactions, (b) identification of invariant moments based on the backward substitution of \req{eq:DM_moment}.}
  \label{Application_example}
\end{figure}

Consider the chemical reaction network shown in \rfig{Application_example}(a), which has an Antithetic Integral Feedback (AIF) structure formed by a sequestration reaction\cite{AIF2016,AIF2019}, and consists of $n = 6$ molecular species and $m = 12$ reactions.
Here, we aim to identify the invariant stationary moments under the variation of $\theta_4$, \textit{i.e.,} $\mathcal{M}_4$. 
Formulating the stationary moment equations up to $\sum^n_{i = 1}\alpha_i = 1$, we have 
\begin{align}
\label{application_ME}
    \!\!\!\!\begin{bmatrix}
0 &\!\! 0 &\!\!\! 0 &\!\!\! 0 &\!\!\! 0 &\!\!\!\! -\theta_{11} &\!\!\!\! 0 \\
\theta_2 &\!\! \theta_3 &\!\!\! 0 &\!\!\! 0 &\!\!\! \theta_9 &\!\!\!\! 0 &\!\!\!\!-\theta_3 \\
0 &\!\! \theta_4 &\!\!\! -\theta_5 &\!\!\! \theta_6 &\!\!\! 0 &\!\!\!\! 0 &\!\!\!\! 0  \\
0 &\!\! 0 &\!\!\! \theta_5 &\!\!\! -\theta_6&\!\!\! 0 &\!\!\!\! 0 &\!\!\!\! 0  \\
0 &\!\! 0 &\!\!\! \theta_7 &\!\!\! 0 &\!\!\! -\theta_8 &\!\!\!\! 0 &\!\!\!\! 0 \\
0 &\!\! 0 &\!\!\! 0 &\!\!\! 0 &\!\!\! \theta_{10} &\!\!\!\! -\theta_{11} &\!\!\!\! 0 \\
\end{bmatrix}\!\!
\begin{bmatrix}
\!\mathbb{E}[x_1]\! \\
\!\mathbb{E}[x_2]\! \\
\!\mathbb{E}[x_3]\! \\
\!\mathbb{E}[x_4]\! \\
\!\mathbb{E}[x_5]\! \\
\!\mathbb{E}[x_1x_6]\! \\
\!\mathbb{E}[x^2_2]\!
\end{bmatrix}
\!\!
=
\!\!
\begin{bmatrix}
\!-\theta_1\!\\
\!0 \!\\
\!0 \!\\
\!-\theta_{12}\!  \\
\!0\!\\
\!0\!
\end{bmatrix}\!\!\!
\end{align}
In this form, the system is underdetermined with six equations and seven variables. 
Applying Lemma \ref{lemma2} and extracting the welldetermined subequations, we obtain eq.~\eqref{eq:DM_moment}, of which the sparsity pattern is shown Fig.~\ref{Application_example}(b). 
Finally, using Theorem 1, we explore invariant
moments $\mathcal{M}_4$. 
Specifically, a graph $\mathcal{G}(\mathcal{V}, \mathcal{E})$ with 10 nodes with two source nodes at $V_7$ and $V_{10}$ is defined since $b_7 \neq 0$ and $b_{10} \neq 0$. 
Then, the reachable states are examined by tracing all reachable nodes from the sources (see the orange lines in \rfig{Application_example}(b)).
The result shows that $\mathbb{E}[x_1x_6], \mathbb{E}[x_5], \mathbb{E}[x_3],$ and $\mathbb{E}[x_4]$ are invariant moments $\mathcal{M}_4$ with respect to the variation of $\theta_4$.

These results agree with the existing results for the AIF systems \cite{AIF2016,gupta2022}, which guarantee the invariance of $\mathbb{E}[x_5]$ and $\mathbb{E}[x_1x_6]$, while the proposed method enables to further discover the invariance of $\mathbb{E}[x_3]$ and $\mathbb{E}[x_4]$. 
This demonstrates that the proposed method can systematically identify invariant moments with respect to arbitrary parameters using a polynomial-time algorithm, addressing a problem that is generally considered challenging due to the infinite dimensionality of the moment equations.
The identified invariant moments provide deeper insights into the homeostatic behavior of biomolecular systems, advancing both the fundamental understanding of biological robustness and the engineering of robust gene circuits for synthetic biology applications.

\section{Conclusion}
In this paper, we have proposed a systematic method for analyzing the parameter invariance of stationary moments in general chemical reaction networks by leveraging the non-zero structure of the moment equations.  
Specifically, we revealed the hierarchical structure among stationary moments using the Dulmage-Mendelsohn (DM) decomposition, and demonstrated that invariant stationary moments 
can be systematically identified by performing unreachability analysis on a graph based on this structure.
The proposed method provides a computationally efficient framework that offers new insights into the homeostatic behavior in biomolecular systems. 
Future work will focus on increasing 
the number of identifiable invariant moments by incorporating parameter dependencies, as well as extending the framework toward the design and control problems of target moments.

\appendix
\subsection{Proof of Lemma \ref{lemma2}}
\label{app.1}
Let $P$ and $Q$ denote the row and column permutation matrices in eq. \eqref{DM}, respectively. Then, we have 
\begin{align}
\!\!\!
            &\begin{bmatrix}
                P&O\\
                O&P
            \end{bmatrix}
\!\!           
            \begin{bmatrix}
                {A}&{A}_k\\
                O & {A}
            \end{bmatrix}
\!\!
            \begin{bmatrix}
                Q&O\\
                O&Q
            \end{bmatrix}
\!\!=\!\!\begin{bmatrix}                {A}_u&{A}_{ud}&{A}^k_u&{A}^k_{ud}\\
                O&{A}_d&O&{A}^k_d\\
                O&O&{A}_u&{A}_{ud}\\
                O&O&O&{A}_d
            \end{bmatrix}\!\!\!\!
            \label{eq:block-transformation}
        \end{align}
with ${A}^k_{u} \in \mathbb{R}^{(M-N_d)\times(N-N_d)}$ and ${A}^k_{ud} \in \mathbb{R}^{(M-N_d)\times(N_d)}$. 
Permuting the second and third diagonal blocks and defining 
        \begin{align}
            {A}_{u1}:=
            \begin{bmatrix}
                {A}_u&{A}^k_{u}\\
                O&{A}_u
            \end{bmatrix},
            {A}_{u2}:=
            \begin{bmatrix}
                {A}_{ud}\\
                O
            \end{bmatrix},
            {A}_{u3}:=
            \begin{bmatrix}
                {A}^k_{ud}\\
                {A}_{ud}
            \end{bmatrix}\notag
        \end{align}
we have eq. \eqref{eq:Psi}. 
Next, we show the maximality of diagonal blocks. 
Each row and each column of the matrices ${A}_d$ and ${A}_u$ contains at least one non-zero element due to Assumption \ref{assumption2} and $\text{rank}({A}_d) = N_d$. 
Therefore, the only admissible row and column permutations in eq. \eqref{eq:block-transformation} that preserve both of the block-triangular and repeated block diagonal structures is (i) swapping of entries within the upper (or lower) half of the matrix or (ii) swapping of identical diagonal blocks that do not change the resulting matrix. 
However, by Lemma \ref{lemma1}, the operation (i) does not increase the dimension of ${A}_d$. Thus, the matrix \eqref{eq:block-transformation} has the maximal diagonal blocks, and their uniqueness follows from Lemma \ref{lemma1}.
\end{document}